\documentclass[twocolumn,nofootinbib]{revtex4}%

\usepackage{amssymb}
\usepackage{amsfonts}
\usepackage{amsmath}
\usepackage{graphicx,color}
\usepackage{slashed}
\usepackage{young}
\usepackage[vcentermath]{youngtab}
\usepackage[utf8]{inputenc}

\usepackage{tensor}

\makeatletter

\newcommand{\be}{\begin{equation}}
\newcommand{\ee}{\end{equation}}
\newcommand{\bea}{\begin{eqnarray}}
\newcommand{\eea}{\end{eqnarray}}

\begin{document}

\title{Chiral Tensors of Mixed Young Symmetry}
\author{Marc Henneaux,  Victor Lekeu, Amaury Leonard}
\affiliation{Universit\'e Libre de Bruxelles and International Solvay Institutes, ULB-Campus Plaine CP231, B-1050 Brussels, Belgium}

\begin{abstract}
Chiral tensors of mixed Young symmetry, which exist in  the same spacetime dimensions  $2 + 4n$ where chiral $p$-forms can be defined, are investigated.   Such chiral tensors have been argued to play a central role in exotic formulations of gravity in 6 dimensions and possess intriguing properties. A variational principle that yields the chiral equations of motion is explicitly constructed and related to the action for a non-chiral tensor. The use of prepotentials turns out to be essential in our analysis.  We also comment on dimensional reduction.
\end{abstract}
\maketitle

\section{Introduction}
\setcounter{equation}{0}

Chiral $p$-forms are an essential building block of various supergravity models \cite{supergravity}.  These forms are defined to be such that their curvature $F$ is self-dual, $F = \! ^* F$ \cite{anti}.
In Minkowski space, chiral $p$-forms exist in dimensions $D = 2 + 4n$ ($n = 0,1, 2, \cdots$) with $p = 2n$, since both $F$ and $\!^* F$ must be forms of the same rank and the Hodge duality operation must square to the identity when acting on those $p$-forms, $(^* )^2 = 1$.

Following \cite{Floreanini:1987as}, an action principle that gives directly the chirality condition was constructed in \cite{Henneaux:1988gg}. This action principle not only automatically yields the chirality condition, which does not need to be separately imposed by hand, but it involves solely the $p$-form gauge potential without auxiliary fields, making the dynamics quite transparent. For instance,  in the free case, it is just quadratic in the fields.  One characteristic feature of the action of \cite{Henneaux:1988gg} is that it is not manifestly covariant in the sense that it {\em is} covariant, but that the fields do not transform off-shell  in the standard way under Lorentz transformations.   When couplings to gravity are included, this means that the fields do not transform in the standard way under spacetime diffeomorphisms.  A similar feature (non-standard transformations under spacetime diffeomorphisms) was encountered in the variational principle recently proposed in \cite{Sen:2015nph}.  

Tensor fields with mixed Young symmetry naturally appear in higher spacetime dimensions in the dual formulation of (linearized) gravity and in general studies of higher spin fields.  They also appear in string field theory where they become massless in the tensionless limit.  

In the same spacetime dimensions $D= 2 + 2p$ ($p$ even) where chiral $p$ forms exist, one can also consistently impose chirality conditions on the curvature tensor of mixed Young symmetry tensors when these are described by Young tableaux, the length of the first column of which is equal to $p$.  This will be explained below. The $p$-form case corresponds to a Young tableau with a single column, the generalization considered here may involve an arbitrary number of columns.

Chirality conditions on mixed Young symmetry tensors have been actually considered in dimension six in the insightful and intriguing work \cite{Hull:2000zn,Hull:2000rr,Hull:2001iu}, where it was argued that the strong coupling limit of theories having $N=8$ supergravity as their low energy effective theory in five spacetime dimensions should be a six-dimensional theory involving, besides chiral $2$-forms, a chiral mixed $(2,2)$-tensor in place of the standard graviton.   In $(2,2)$,  the numbers give the number of boxes in the successive rows of the corresponding Young tableau. A very attractive feature of the six-dimensional theory containing the $(2,2)$ chiral tensor is that it provides a remarkable geometric interpretation of electric-magnetic gravitational duality in four dimensions.

Now, the discussion of \cite{Hull:2000zn,Hull:2000rr,Hull:2001iu} was performed at the level of the equations of motion.  Although one may develop quantization methods that bypass the Lagrangian, the investigation of the quantum properties lies definitely on more familiar grounds when a self-contained action principle does exist.

The central result of our paper is to establish the existence of a local variational principle for free chiral tensors of mixed symmetry,   which we explicitly write down.   To achieve this goal,  one must write the equations of motion in terms  of prepotentials generalizing those introduced in \cite{Henneaux:2004jw}.  Our approach relies on the tools for dealing with prepotentials and duality developed in \cite{Henneaux:2015cda,Henneaux:2016zlu}.
As it is the case for chiral $p$-forms, the action is not  manifestly Lorentz-invariant, even though the space of solutions is. 

\section{Chiral $(2,2)$-tensor in $D=6$ spacetime dimensions}
\setcounter{equation}{0}

\subsection{Equations of motion}
To illustrate the features brought in by tensors of mixed symmetry type, without having to deal with the extra technical complications of trace conditions, we concentrate on the simplest case, namely, tensors with $(2,2)$ Young symmetry in $D=6$ spacetime dimensions. This case is described by the Young tableau
{\tiny $\yng(2,2).$}
The general case will be discussed at the end of this letter.   We denote the corresponding gauge field by $T_{\alpha_1 \alpha_2 \beta_1 \beta_2}$, with $T_{\alpha_1 \alpha_2 \beta_1 \beta_2} = - T_{\alpha_2 \alpha_1 \beta_1 \beta_2} = - T_{\alpha_1 \alpha_2 \beta_2 \beta_1}$ and $T_{[\alpha_1 \alpha_2 \beta_1] \beta_2} =0$.  The gauge symmetries are $\delta T_{\alpha_1 \alpha_2 \beta_1 \beta_2}= \mathbb{P}_{(2,2)} \left( \partial_{\alpha_1} \eta_{ \beta_1 \beta_2 \alpha_2} \right)$ where $\eta_{ \beta_1 \beta_2 \alpha_2} $ is an arbitrary $(2,1)$-tensor.  Here, $\mathbb{P}_{(2,2)}$ is the projector on the $(2,2)$ symmetry.

The gauge invariant curvature, or ``Riemann tensor",  is a tensor of type $(2,2,2)$,
$R\sim  $ {\tiny $  \yng(2,2,2)$ }. In components , $ R_{\alpha_1 \alpha_2 \alpha_3 \beta_1 \beta_2 \beta_3} = \partial_{[\alpha_1}T_{\alpha_2 \alpha_3][ \beta_1 \beta_2, \beta_3]}$ and one has $R_{\alpha_1 \alpha_2 \alpha_3 \beta_1 \beta_2 \beta_3}= R_{[\alpha_1 \alpha_2 \alpha_3] \beta_1 \beta_2 \beta_3}= R_{\alpha_1 \alpha_2 \alpha_3 [\beta_1 \beta_2 \beta_3]}$ as well as \begin{equation} R_{[\alpha_1 \alpha_2 \alpha_3 \beta_1] \beta_2 \beta_3} = 0. \label{cyclic} \end{equation} 

The equations of motion for a general $(2,2)$-tensor express that the corresponding $(2,2)$ ``Ricci tensor", i.e., the trace 
$ R_{\alpha_1 \alpha_2  \beta_1 \beta_2 } \equiv  R_{\alpha_1 \alpha_2 \alpha_3 \beta_1 \beta_2 \beta_3} \eta^{\alpha_3 \beta_3}$ of the Riemann tensor,  vanishes,
\begin{equation}
 R_{\alpha_1 \alpha_2  \beta_1 \beta_2 } = 0. \label{EOM0}
 \end{equation}
 
 In 6 spacetime dimensions, the dual $^*R$ of the Riemann tensor on, say, the first three indices
 $
 ^*R_{\alpha_1 \alpha_2 \alpha_3 \beta_1 \beta_2 \beta_3} = \frac{1}{3!} \epsilon_{\alpha_1 \alpha_2 \alpha_3 \lambda_1 \lambda_2 \lambda_3} R^{\lambda_1 \lambda_2 \lambda_3}_{\; \; \; \; \; \; \; \; \; \; \; \; \beta_1 \beta_2 \beta_3} 
 $
 is traceless because of the cyclic identity (\ref{cyclic}), i.e., 
 \be
 ^*R_{\alpha_1 \alpha_2  \beta_1 \beta_2 } = 0.
\ee
 This implies that a $(2,2)$-tensor field $T$ with a self-dual or anti-self-dual Riemann tensor
 \begin{equation}
 R =  \;   \!^*R \label{SD22}
 \end{equation}
 (self-duality) or $R = - ^*R$ (anti-self-duality)
 is automatically a solution of the equations of motion (\ref{EOM0}).   Note that this implies that $^*R$ is also a $(2,2,2)$ tensor.  
 The condition (\ref{SD22}) is consistent because $(^*)^2 = 1$ in this case.  
 The question addressed in this note is to derive (\ref{SD22}) from a variational principle.  

There is a mismatch between the number of equations (\ref{SD22}), namely $175$, and the number of components of the $(2,2)$-tensor field,  namely $105$.  But the equations (\ref{SD22}) are not all independent.  It is of course sufficient that the searched-for variational principle yields a system of equations equivalent to (\ref{SD22}).
 
 \subsection{Electric and magnetic fields}
 
 To identify such a subset derivable from a variational principle, we introduce the electric and magnetic fields.  The electric field contains the components of the curvature tensor with the maximum number of indices equal to the time direction $0$, namely, two, 
 ${\mathcal E}^{ijkl}  \sim R^{0ij0kl} $,  or what is the same on-shell, the components of the curvature with no index equal to zero.  Since in 5 dimensions, the curvature tensor $R_{pijqkl}$ is completely determined by the Einstein tensor $G\indices{^{ij}_{kl}} = \frac{1}{(3!)^2} R\indices{^{abc def}} \varepsilon\indices{_{abc}^{ij}} \varepsilon_{defkl} = R\indices{^{ij}_{kl}} - 2 \delta^{[i}_{[k} R\indices{^{j]}_{l]}} + \frac{1}{3} \delta^i_{[k} \delta^j_{l]} R$ (the Weyl tensor identically vanishes),  one defines explicitly the electric field as
 \be {\mathcal E}^{ijkl} \equiv G^{ijkl}. \ee
 Here, $R\indices{^{ij}_{kl}} = R\indices{^{mij}_{mkl}} $, $R\indices{^{j}_{l}} = R\indices{^{mij}_{mil}}$ and 
$R = R\indices{^{mij}_{mij}} $ are the successive traces.  Similar conventions will be adopted below for the traces of the tensors that appear.
 The electric field has the $(2,2)$ Young symmetry and is identically transverse, $ \partial_i {\mathcal E}^{ijkl} = 0$.  It is also traceless on-shell,
 \be
 {\mathcal E}^{ik} \equiv {\mathcal E}^{ijkl}\delta_{jl} = 0 . \label{trace0}
 \ee
 
 The magnetic field contains the components of the curvature tensor with only one index equal to $0$,
 \be
 {\mathcal B}_{ijkl}  = \frac{1}{3!} R\indices{_{0ij}^{abc}} \varepsilon_{abckl} .
 \ee
It is identically traceless, ${\mathcal B}^{jl} \equiv {\mathcal B}^{ijkl} \delta_{ik} = 0$, and transverse on the second pair of indices, $ \partial_k {\mathcal B}^{ijkl} = 0$.  On-shell, it has the $(2,2)$ Young symmetry.
 
The self-duality equation (\ref{SD22}) implies
\be
{\mathcal E}^{ijrs} - {\mathcal B}^{ijrs} = 0 . \label{E=B}
\ee
Conversely, the equation (\ref{E=B}) implies all the components of the self-duality equation (\ref{SD22}). This is verified in appendix \ref{app:eom} by repeating the argument of \cite{Bunster:2012km} given there for a $(2)$-tensor, which is easily adapted to a $(2,2)$-tensor. 
We have thus replaced the self-duality conditions (\ref{SD22}) by a smaller, equivalent, subset.  One central feature of this subset is that it is expressed in terms of spatial objects.  

Note that the trace condition (\ref{trace0}) directly follows by taking the trace of (\ref{E=B}) since the magnetic field is traceless. It appears as a constraint on the initial conditions because it does not involve the time derivatives of $T_{ijrs}$.  There is no analogous constraint in the $p$-form case.

Since the number of components of the electric field is equal to the number of spatial components $T_{ijrs}$ of the $(2,2)$-tensor $T_{\alpha \beta \lambda \mu}$, one might wonder whether the equations (\ref{E=B}) can be derived from an action principle in which the basic variables would be the $T_{ijrs}$.   This does not work, however. 
Indeed, while the electric field involves only the spatial components $T_{ijrs}$ of the gauge field, the magnetic field involves also the gauge component $T_{0jrs}$, through an exterior derivative.  One must therefore get rid of $T_{0jrs}$. 

To get  equations that involve only the spatial components $T_{ijrs}$, we proceed as in the $2$-form case and take the curl of (\ref{E=B}), i.e.
\be
\epsilon^{mnijk}\partial_{k}\left({\mathcal E}_{ij}^{\; \; \; \; rs} - {\mathcal B}_{ij}^{\; \; \; \; rs} \right)=0 , \label{E2=B2}
\ee
eliminating thereby  the gauge components $T_{0jrs}$.   We also retain the equation (\ref{trace0}), which is a consequence of (\ref{E=B}) involving only the electric field.  There is no loss of physical information in going from (\ref{E=B}) to the system (\ref{trace0}), (\ref{E2=B2}). Indeed, as shown in appendix \ref{app:eom}, if (\ref{trace0}) and (\ref{E2=B2}) are fulfilled, one recovers (\ref{E=B}) up to a term that can be absorbed in a redefinition of $T_{0jrs}$.  The use of (\ref{trace0}) is crucial in the argument.  It is in the form (\ref{trace0}), (\ref{E2=B2}) that the self-duality equations can be derived from a variational principle.

\subsection{Prepotentials - Action}

To achieve the goal of constructing the action for the chiral tensor, we first solve the constraint (\ref{trace0}) by introducing a prepotential $Z_{ijrs}$ for $T_{ijrs}$.   Prepotentials were defined for gravity in \cite{Henneaux:1988gg} and generalized to arbitrary symmetric tensor gauge fields in \cite{Henneaux:2015cda,Henneaux:2016zlu}. 
The introduction of a prepotential for the mixed tensor $T_{ijrs}$ proceeds along similar lines. 

Explicitly, the prepotential $Z_{ijrs}$ provides a parametrization of the most general $(2,2)$ tensor field $T_{ijrs}$ that solves the constraint (\ref{trace0}). One has
\be 
T_{ijrs} = {\mathbb P}_{(2,2)} \left( \frac{1}{3!} \epsilon_{ij}^{\; \; \;   k mn} \partial_k Z_{mn rs} \right) + \hbox{gauge transf.}, \label{TZ}
\ee
which is a direct generalisation of the formula given in \cite{Henneaux:2004jw} for a $(2)$-tensor.   The prepotential is determined up to the gauge symmetries
\be
\delta Z_{ijrs}  = {\mathbb P}_{(2,2)} \left(\partial_{i} \xi_{rs j} + \lambda_{ir}\delta_{js}\right) \label{Weyl}
\ee
where $\xi_{rs j}$ is a $(2,1)$-tensor parametrizing the ``linearized spin-$(2,2)$ diffeomorphisms" of the prepotential and $\lambda_{ir}$ a symmetric tensor parametrizing its ``linearized spin-$(2,2)$ Weyl rescalings".

Because the Weyl tensor of a $(2,2)$-tensor identically vanishes, the relevant tensor that controls Weyl invariance is the ``Cotton tensor", defined as
\be D_{ij kl} = \frac{1}{3!} \varepsilon_{ijabc} \partial^{a} S\indices{^{bc}_{kl}},
\ee
where $S\indices{^{ij}_{kl}} = G\indices{^{ij}_{kl}} - 2 \delta^{[i}_{[k} G\indices{^{j]}_{l]}} + \frac{1}{3} \delta^i_{[k} \delta^j_{l]} G$ is the ``Schouten tensor", which has the key property of transforming as $\delta S\indices{^{ij}_{kl}} = -\frac{4}{27} \partial^{[j}\partial_{[k} \lambda\indices{^{i]}_{l]}}$ under Weyl rescalings. 
The Cotton tensor $D_{ij kl}$  is a $(2,2)$-tensor which is gauge invariant under (\ref{Weyl}), as well as identically transverse and traceless, $\partial_i D^{ij rs} = 0 = D^{ij rs}  \delta_{js}$.  Furthermore, a necessary and sufficient condition for $Z_{ijrs}$ to be pure gauge is that its Cotton tensor vanishes.

The relation (\ref{TZ}) implies that 
\be
\mathcal{E}^{ij rs} [T[Z]] \equiv G^{ij rs} [T[Z]] = D^{ij rs} [Z].
\ee
The relation (\ref{TZ}) gives the most general solution for $T_{ij rs}$ subject to the constraint that $\mathcal{E}^{ij rs}$ is traceless (this is proved in   \cite{Henneaux:2015cda} for general higher spins described by completely symmetric tensors, and is easily extended to tensors with mixed Young symmetry).  We note that in three dimensions, the analogous relations on the Cotton tensor for symmetric gauge fields have a nice supersymmetric interpretation \cite{Kuzenko}. It would be of interest to explore whether a similar interpretation holds here.

It follows from (\ref{TZ}) that 
\be
\frac12 \epsilon^{mnijk}\partial_{k}{\mathcal B}_{ij}^{\; \; \; \; rs}= \dot{D}^{mn rs} [Z]
\ee
and therefore, in terms of the prepotential $Z_{ijrs}$, the self-duality condition (\ref{E2=B2}) reads
\be
\frac12 \epsilon^{mnijk}\partial_{k}D_{ij}^{\; \; \; \; rs}[Z]  - \dot{D}^{mn rs} [Z]=0,
\ee
an equation that we can rewrite as
\be L^{mnrs \vert ijpq} Z_{ijpq} = 0 \label{E3=B3}
\ee
where the differential operator $L^{mnrs \vert ijpq}$ contains four derivatives and can easily be read off from (\ref{E3=B3}).  The operator $L^{mnrs \vert ijpq}$ is {\em symmetric}, so that one can form the action 
\begin{eqnarray} 
&& S[Z] = \frac12 \int d^6x  Z_{mnrs} \left(L^{mnrs \vert ijpq} Z_{ijpq}\right) \label{Action} \\
&&= \frac12 \int d^6x  Z_{mnrs} \left(\dot{D}^{mn rs} [Z] - \frac12\epsilon^{mnijk}\partial_{k}D_{ij}^{\; \; \; \; rs}[Z]\right)  \nonumber
\end{eqnarray}
which yields (\ref{E3=B3}) as equations of motion.  
Given that $Z \sim \partial^{-1} T$,  this action contains the correct number of derivatives of $T$, namely two,  and has therefore the correct dimension.

\subsection{Chiral and non-chiral actions}

The action (\ref{Action}) is our central result.  Although not manifestly so, it is  covariant.  One way to see this is to observe that (\ref{Action}) can be derived from the manifestly covariant Curtright action for a $(2,2)$-field \cite{Curtright:1980yk,Boulanger:2004rx} rewritten in Hamiltonian form.  As explained in appendix \ref{app:ham} , this action involves the spatial components $T_{ijrs}$ and their conjugate momenta $\pi^{ijrs}$ as canonically conjugate dynamical variables, while the temporal components $T_{0ijk}$ and $T_{0i0j}$ play the role of Lagrange multipliers for the ``momentum constraint"
$\mathcal{C}^{ijk} \equiv \partial_l \pi^{ijlk} \approx 0$ and the ``Hamiltonian constraint" $\mathcal{C}^{ij} \equiv \mathcal{E}\indices{^{ikj}_k}[T] \approx 0$.  These constraints can be solved by introducing two prepotentials $Z^{(1)}_{ijrs}$ and $Z^{(2)}_{ijrs}$.   As for a chiral $2$-form \cite{Bekaert:1998yp}, the linear change of variables $(Z^{(1)}_{ijrs}, Z^{(2)}_{ijrs}) \rightarrow (Z^+_{ijrs}= Z^{(1)}_{ijrs} + Z^{(2)}_{ijrs}, Z^-_{ijrs} = Z^{(1)}_{ijrs} - Z^{(2)}_{ijrs})$ splits the action as a sum of two independent terms, one for $Z^+_{ijrs}$ and one for $Z^-_{ijrs}$.   The Poincar\'e generators also split similarly,   one for $Z^+_{ijrs}$ and one for $Z^-_{ijrs}$, which transform separately.  The action (\ref{Action}) is the action for $Z^+_{ijrs}$ obtained though this decomposition procedure, with the identification $Z^+_{ijrs} \equiv Z_{ijrs}$.  This second method for obtaining the action for a chiral $(2,2)$ tensor  shows as a bonus how a  non-chiral $(2,2)$-tensor dynamically splits as the sum of a chiral $(2,2)$-tensor  and an  anti-chiral $(2,2)$-tensor.

\subsection{Dimensional reduction}

Upon reduction from $5+1$ to $4+1$ dimensions, the prepotential $Z_{ijrs}$ decomposes into a $(2,2)$-tensor, a $(2,1)$-tensor and a $(2)$-tensor.  Using part of the Weyl symmetry, one can set the $(2)$-tensor equal to zero, leaving  one with a  $(2,2)$-tensor and a $(2,1)$-tensor which are exactly the prepotentials of the pure Pauli-Fierz theory in $4+1$ dimensions \cite{Bunster:2013oaa}, with the same action and gauge symmetries (see appendix \ref{app:dimred}).  It is this remarkable connection between the $(2,2)$-self-dual theory in $6$ spacetime dimensions and pure (linearized) gravity in $5$ spacetime dimensions that is at the heart of the work \cite{Hull:2000zn,Hull:2000rr}.  We have shown here that the connection holds not just for the equations of motion, but also for the actions themselves.  

\section{Generalizations and conclusions}

The extension to more general two-column Young symmetry tensors is direct. The ``critical dimensions" where one can impose self-duality conditions on the curvature are those where chiral $p$-forms exist. The first colum of the Young tableau characterizing the Young symmetry must have $p$ boxes, and the second column has then a number $q \leq p$ of boxes.  So, in $D=6$ spacetime dimensions, one has also the interesting case of $(2,1)$-tensors, also considered in \cite{Hull:2000zn,Hull:2000rr}.  This case is treated along lines identical to those described here.  For the next case -- $D=10$ spacetime dimensions --, the first column must have length 4, and the second colum has length $q \leq 4$, an interesting example being the  $(2,2,2,2)$-tensors.  Again, the extension to this two-column symmetry case is direct, as in all higher spacetime dimensions $D= 14, 18, 22, 26 , \cdots$.  

The extension to more than two column Young symmetries is more subtle but proceeds as in \cite{Henneaux:2016zlu}, by relying on the crucial property demonstrated in \cite{Bekaert:2003az}, where it was shown that the second-order Fronsdal-Crurtright type equations can be replaced by equations on the curvatures, which involves higher order derivatives.  The self-duality conditions can then be derived from an action principle involving the appropriate prepotentials.  The action is obtained by combining the above derivation with the methods of  \cite{Henneaux:2016zlu} for introducing prepotentials.

The present analysis can be developed in various directions.  First, following  \cite{Hull:2000zn,Hull:2000rr}, it would be of great interest to consider the supersymmetric extensions of the $6$-dimensional chiral theory and to determine how the fermionic prepotentials enter the picture \cite{susy22}. The attractive $(4,0)$-theory of \cite{Hull:2000zn,Hull:2000rr} deserves a particular effort in this respect \cite{Future}.  Second, the inclusion of sources, which would be dyonic by the self-duality condition, and the study of the corresponding quantization conditions, would also be worth understanding \cite{Deser,Seiberg:2011dr,Bunster:2013era}. 

Finally, we note that we restricted the analysis to flat Minkowski space.  The trivial topology of $\mathbb{R}^n$ enabled us to integrate the differential equations for the prepotentials without encountering obstructions, using the Poincar\'e lemma of \cite{DuboisViolette}. The consideration of Minkowski space is not optional at this stage since the coupling of a single higher spin field to curved backgrounds is problematic.  It is known how to surpass the problems only in the context of the Vasiliev theory, which requires an infinite number of fields \cite{Vasiliev:1995dn,Vasiliev:2004cp,Bekaert:2010hw,Didenko:2014dwa,Metsaev:1993mj}.  Important ingredients to extend the analysis of the present article to nonlinear backgrounds are expected to include the cohomological considerations of \cite{Bekaert:1998yp}, the nonlinear extension of the higher spin Cotton tensors \cite{Linander:2016brv}, as well as duality in cosmological backgrounds \cite{Julia:2005ze,Hortner:2016omi}.

\section*{Acknowledgments} 
 We thank Xavier Bekaert and Andrea Campoleoni for useful discussions. V. L. and A.L. are Research Fellows at the Belgian F.R.S.-FNRS. This work was partially supported by the ERC Advanced Grant ``High-Spin-Grav", by FNRS-Belgium (convention FRFC PDR T.1025.14 and  convention IISN 4.4503.15) and by the ``Communaut\'e Fran\c{c}aise de Belgique" through the ARC program. 
 \break

\begin{appendix}

\section{Equations of motion} \label{app:eom}

In this appendix, we show the equivalences between the different forms of the self-duality equations given in the main text., \eqref{SD22} $\Leftrightarrow$ \eqref{E=B} $\Leftrightarrow$ \eqref{trace0},\eqref{E2=B2}.

\eqref{SD22} $\Leftrightarrow$ \eqref{E=B}:
In components, the self-duality equation $R = \,^*\! R$ reads
\begin{align}
R_{0ijklm} &= \frac{1}{3!} \varepsilon\indices{_{ij}^{abc}}  R_{abcklm} \label{R0} \\
R_{0ij0kl} &= \frac{1}{3!} \varepsilon\indices{_{ij}^{abc}}  R_{abc0kl} \label{R00}.
\end{align}
The first of these equations is equivalent to \eqref{E=B} by dualizing on the $klm$ indices. Conversely,  we must show that \eqref{E=B} implies \eqref{R00} or, equivalently, that \eqref{R0} implies \eqref{R00}. To do so, we use the Bianchi identity $\partial_{[\alpha_1} R_{\alpha_2\alpha_3\alpha_4]\beta_1 \beta_2 \beta_3} = 0$ on the curvature, which imples
\begin{equation}
\partial_0 R_{ijk \beta_1 \beta_2 \beta_3} = 3 \partial_{[i} R_{jk]0 \beta_1 \beta_2 \beta_3} .
\end{equation}
Therefore, taking the time derivative of equation \eqref{R0} gives
\begin{equation}
\partial_{[k} R_{lm]00ij} = \frac{1}{3!} \partial_{[k} R_{lm]0 abc} \varepsilon\indices{_{ij}^{abc}}, \label{curlR00}
\end{equation}
which is exactly the curl of \eqref{R00}. Now, the tensor $R_{0lm0ij}$ has the $(2,2)$ symmetry, and so does $\frac{1}{3!} R_{0lm abc} \varepsilon\indices{_{ij}^{abc}} = \mathcal{B}_{lmij}$ because of equation \eqref{E=B} and the fact that $\mathcal{E}$ has the $(2,2)$ symmetry. Using the Poincaré lemma of \cite{DuboisViolette} for rectangular Young tableaux, one recovers equation \eqref{R00} up to a term of the form $\partial_{[i} N_{j][k,l]}$ for $N_{jk}$ symmetric. This term can be absorbed in a redefinition of the $T_{0j0k}$ components appearing in $R_{0ij0kl}$. (In fact, the components $T_{0j0k}$ drop from equation \eqref{curlR00}, and this explains how one can get equation \eqref{R00} from \eqref{R0}, which does not contain $T_{0j0k}$ either.)

\eqref{E=B} $\Leftrightarrow$ \eqref{trace0},\eqref{E2=B2}:
Equation \eqref{E=B} obviously implies \eqref{E2=B2}. It also implies \eqref{trace0} because the magnetic field $\mathcal{B}$ is identically traceless. To prove the converse, we introduce the tensor $K_{ijk lm} = \varepsilon\indices{_{ijk}^{ab}} (\mathcal{E} - \mathcal{B})_{lmab}$. Equation \eqref{trace0} and the fact that $\mathcal{B}$ is traceless imply that $K$ has the $(2,2,1)$ symmetry, $K\sim  $ {\tiny $  \yng(2,2,1)$ }. Equation \eqref{E2=B2} states that the curl of $K$ on its second group of indices vanishes, $K_{ijk[lm,n]} = 0$. The explicit formula
\begin{equation}
K_{ijk lm} = \frac{1}{3} \left(\varepsilon_{lmpqr} \partial^{p} T\indices{^{qr}_{[ij,k]}} - \partial_{[0} T_{lm][ij,k]} \right)
\end{equation}
shows that the curl of $K$ on its first group of indices also vanishes, $\partial_{[i} K_{jkl]mn} = 0$. Using the generalized Poincaré lemma of \cite{Bekaert:2002dt} for arbitrary Young tableaux, this implies that $K_{ijklm} = \partial_{[i} \lambda_{jk][l,m]}$, where $\lambda_{jkl}$ is a tensor with the $(2,1)$ symmetry that can be absorbed in a redefinition of $T_{0ijk}$. (Similarly to the previous case, those components actually drop from \eqref{E2=B2}.) One finally recovers equation \eqref{E=B} by dualizing again $K$ on its first group of indices.

\section{Hamiltonian formulation} \label{app:ham}

The Lagrangian for a non-chiral $(2,2)$ tensor $T_{\mu\nu\rho\sigma}$ is given by \cite{Boulanger:2004rx}
\begin{equation}
\mathcal{L} = - \frac{5}{2} \,\delta^{\mu_1 \dots \mu_5}_{\nu_1 \dots \nu_5} \, M\indices{^{\nu_1\nu_2\nu_3}_{\mu_1\mu_2}} \, M\indices{_{\mu_3\mu_4\mu_5}^{\nu_4\nu_5}},
\end{equation}
where $M_{\mu\nu\rho \sigma\tau} = \partial_{[\mu}T_{\nu\rho]\sigma\tau}$ and $\delta^{\mu_1 \dots \mu_5}_{\nu_1 \dots \nu_5} = \delta^{\mu_1}_{[\nu_1} \dotsm \delta^{\mu_5}_{\nu_5]}$.
The associated Hamiltonian action is
\begin{equation}
S_H = \int \! dt \, d^5\! x \left( \pi_{ijkl} \dot{T}^{ijkl} - \mathcal{H} - n_{ijk} \,\mathcal{C}^{ijk} - n_{ij}\, \mathcal{C}^{ij} \right),
\end{equation}
where the Hamiltonian is
\begin{align}
\mathcal{H} &= \mathcal{H}_\pi + \mathcal{H}_T \\
\mathcal{H}_\pi &= 3 \left( \pi^{ijkl}\pi_{ijkl} - 2 \pi^{ij}\pi_{ij} + \frac{1}{3} \pi^2\right) \\
\mathcal{H}_T &= \frac{5}{2}\, \delta^{i_1 \dots i_5}_{j_1 \dots j_5} \, M\indices{^{j_1 j_2 j_3}_{i_1 i_2}} \, M\indices{_{i_3 i_4 i_5}^{j_4 j_5}} .
\end{align}
The components $n_{ijk} = - 4 T_{ij0k}$ and $n_{ij} = 6 T_{0i0j}$ of $T$ with some indices equal to zero only appear as Lagrange multipliers for the constraints
\begin{align}
\mathcal{C}^{ijk} &\equiv \partial_l \pi^{ijlk} = 0\\
\mathcal{C}^{ij} &\equiv \mathcal{E}\indices{^{ikj}_k}[T] = 0.
\end{align}
Those constraints are solved by introducing two prepotentials $Z^{(1)}_{ijkl}$ and $Z^{(2)}_{ijkl}$ through
\begin{align}
\pi^{ijkl} &= G^{ijkl}[Z^{(1)}] \\
T_{ijkl} &= \frac{1}{3} \,{\mathbb P}_{(2,2)} \left( \epsilon\indices{_{ij}^{abc}} \partial_a Z^{(2)}_{bckl} \right) .
\end{align}
In terms of prepotentials, we have up to a total derivative
\begin{align}
\pi_{ijkl} \dot{T}^{ijkl} &= 2\, Z^{(1)}_{ijkl} \dot{D}^{ijkl}[Z^{(2)}] \\
\mathcal{H}_\pi &= 3\, G_{ijkl}[Z^{(1)}]S^{ijkl}[Z^{(1)}] \\ \mathcal{H}_T &= 3\, G_{ijkl}[Z^{(2)}]S^{ijkl}[Z^{(2)}] .
\end{align}
Again up to a total derivative, one has $G_{ijkl}[Z]S^{ijkl}[Z] = \frac{1}{3!} Z_{ijkl} \epsilon^{ijabc}\partial_a D\indices{_{bc}^{kl}}[Z]$. Therefore, defining the prepotentials $Z^{\pm}_{ijkl} = Z^{(1)}_{ijkl} \pm Z^{(2)}_{ijkl}$, the action splits into two parts, $S[Z^+, Z^-] = S^+[Z^+] - S^-[Z^-]$. The action $S^+[Z^+]$ is exactly the action \eqref{Action} provided in the text for a chiral tensor, while $S^-[Z^-]$ is the analog action for an anti-chiral tensor (which differs from equation \eqref{Action} only by the sign of the second term).

\section{Dimensional reduction} \label{app:dimred}

The prepotential decomposes into three tensors,
\begin{equation}
Z_{IJKL} \longrightarrow Z_{ijkl},\; Z_{ij k5},\; Z_{i5j5} .
\end{equation}
(For the purposes of this appendix, uppercase indices run from $1$ to $5$ while lowercase indices run from $1$ to $4$.)
The $(2,2)$-tensor and the $(2,1)$-tensor are identified with the two prepotentials $P_{ijkl}$ and $\Phi_{ijk}$ for linearized gravity in $4+1$ dimensions \cite{Bunster:2013oaa} as
\begin{equation}
Z_{ijkl} = 12 \sqrt{3} P_{ijkl}, \qquad Z_{ijk5} = -3 \sqrt{3} \Phi_{ijk}.
\end{equation}
The $(2)$-tensor $Z_{i5j5}$ transforms under the Weyl symmetries \eqref{Weyl} as $\delta Z_{i5j5} = \frac{1}{3} ( \lambda_{ij} + \delta_{ij} \lambda_{55} )$ and can therefore be set to zero. Remaining gauge transformations on $Z$ must respect this choice: this restricts the gauge parameters to $\lambda_{ij} = - \delta_{ij} \lambda_{55}$ and $\xi_{i55} = 0$. The surviving gauge parameters are then $\lambda_{55}$, $\lambda_{i5}$, $\xi_{ijk}$ and $\xi_{5ij}$. (Note that $\xi_{ij5} = - 2 \xi_{5[ij]}$ is not independent, due to the cyclic identity $\xi_{[IJK]}=0$.) The map with the gauge parameters of \cite{Bunster:2013oaa} is
\begin{align}
\chi_{ijk}&=-\frac{\xi_{ijk}}{24\sqrt{3}}, \quad S_{ij} = \frac{\xi_{5(ij)}}{12\sqrt{3}}, \quad A_{ij} = -\frac{\xi_{5[ij]}}{12\sqrt{3}} \nonumber \\
\xi &= \frac{2\lambda_{55}}{9\sqrt{3}},\quad B_i = \frac{2\lambda_{i5}}{9\sqrt{3}} .
\end{align}
This shows that field content and gauge symmetries match. For the comparison of the actions, one needs the following expressions for the reduction of the Cotton tensor:
\begin{align}
D\indices{_{ij}^{kl}} &= -\frac{2}{\sqrt{3}} \,\varepsilon_{ijab} \partial^a (E^{klb} + \delta^{b[k} E^{l]} ) \\
D\indices{_{ij}^{k5}} &= 2\sqrt{3} \,\varepsilon_{ijab} \partial^a ( R^{kb} - \frac{1}{3} \delta^{kb} R )\\
D\indices{_{i5}^{j5}} &= \frac{1}{\sqrt{3}} \,\varepsilon_{iabc} \partial^a ( E^{bcj} + \delta^{jb} E^c ) ,
\end{align}
where $R^{ij}[P]$ and $E^{ijk}[\Phi]$ are defined as
\begin{align}
R^{ij}[P] &= \frac{1}{(3!)^2} \varepsilon^{iabc} \varepsilon^{jdef} \partial_a \partial_f P_{bcde} \\
E^{ijk}[\Phi] &= \frac{1}{2.3!} \varepsilon^{ij de} \varepsilon^{k abc} \partial_a \partial_e \Phi_{bcd}
\end{align}
and the traces are $R = R\indices{^i_i}$, $E^i=E\indices{^{ij}_j}$.
Using these formulas, one recovers the action of \cite{Bunster:2013oaa} for linearized gravity in $4+1$ dimensions in the prepotential formalism.

\end{appendix}


\begin{thebibliography}{99}
\bibitem{anti}  Anti chiral $p$-forms are such as $F = - \, \! ^* F$ and enjoy similar properties.  For definiteness, we shall focus here on the chiral case.
\end{thebibliography}
\end{document}